\documentclass{article}

\usepackage[margin=2.5cm]{geometry}
\usepackage{lineno,hyperref}
\usepackage{changepage}
\usepackage[utf8x]{inputenc}
\usepackage{graphicx}
\usepackage{setspace} 
\usepackage{textcomp}

\usepackage{natbib}
\setcitestyle{numbers}

\begin{document}

\vspace*{0.2in}

\begin{flushleft}
{\Large
\textbf\newline{\textbf{Towards learning-to-learn}}
}
\newline
\\
Benjamin James Lansdell\textsuperscript{1*},
Konrad Paul Kording\textsuperscript{1},
\\
\bigskip
\textbf{\textsuperscript{1}} Department of Bioengineering, University of Pennsylvania, Philadelphia, PA, USA
\\
\bigskip
* lansdell@seas.upenn.edu

\end{flushleft}

\section*{Abstract}
In good old-fashioned artificial intelligence (GOFAI), humans specified systems that solved problems. Much of the recent progress in AI has come from replacing human insights by learning. However, learning itself is still usually built by humans -- specifically the choice that parameter updates should follow the gradient of a cost function. Yet, in analogy with GOFAI, there is no reason to believe that humans are particularly good at defining such learning systems: we may expect learning itself to be better if we learn it. Recent research in machine learning has started to realize the benefits of that strategy. We should thus expect this to be relevant for neuroscience: how could the correct learning rules be acquired? Indeed, cognitive science has long shown that humans learn-to-learn, which is potentially responsible for their impressive learning abilities. Here we discuss ideas across machine learning, neuroscience, and cognitive science that matter for the principle of learning-to-learn.

\section{Introduction}

Many approaches in artificial intelligence (AI) have relied on hand-crafting solutions for specific tasks. This follows the tradition of what is called good old-fashioned AI \cite{Clark1988-lf}, which sought to explicitly implement our own reasoning processes and knowledge to solve a given problem -- from codifying mathematical reasoning \cite{Newell1956-ay} to expert knowledge in technical domains \cite{Jackson1990-qk}. While this leads to some success, ultimately hand-coding solutions is rigid and difficult to scale to complicated problems. A separate approach, originating in the connectionist and parallel distributed processing schools \cite{Rumelhart1987-ks}, instead focused on using artificial neural networks to solve problems without explicitly coding representations or solution procedures. Recently, the potential of neural networks to solve challenging problems has begun to be realized: deep neural networks can learn directly from data how to solve a given problem, surpassing both hand-crafted algorithms and human-level performance in a number of domains \cite{Mnih2015-io,Silver2017-hp,LeCun2015-yo,He2015-oe,Haenssle2018-nj,Russakovsky2015-hw}. Learning is the centerpiece of modern AI.

However current methods based on deep learning are data inefficient \cite{Lake2015-qt,Tsividis2017-bg}. While a neural network may learn to classify images better than humans, it requires countless labeled images to do so \cite{He2015-oe}. Humans, by contrast, can successfully identify novel categories of objects after seeing only a handful of instances, what is called one-shot or few-shot learning \cite{Lake2017-gi}. One reason for this inefficiency is that neural networks do not generalize at a human level. They can not meaningfully exploit similarities across previously learned tasks. Recent reviews \cite{Lake2017-gi,Marblestone2016-wx,Hassabis2017-hu} have argued that efficient learning is important to realizing human-like learning in AI. 

\section{Learning-to-learn in AI}

\begin{figure}
  \centering
      \includegraphics[width=0.8\textwidth]{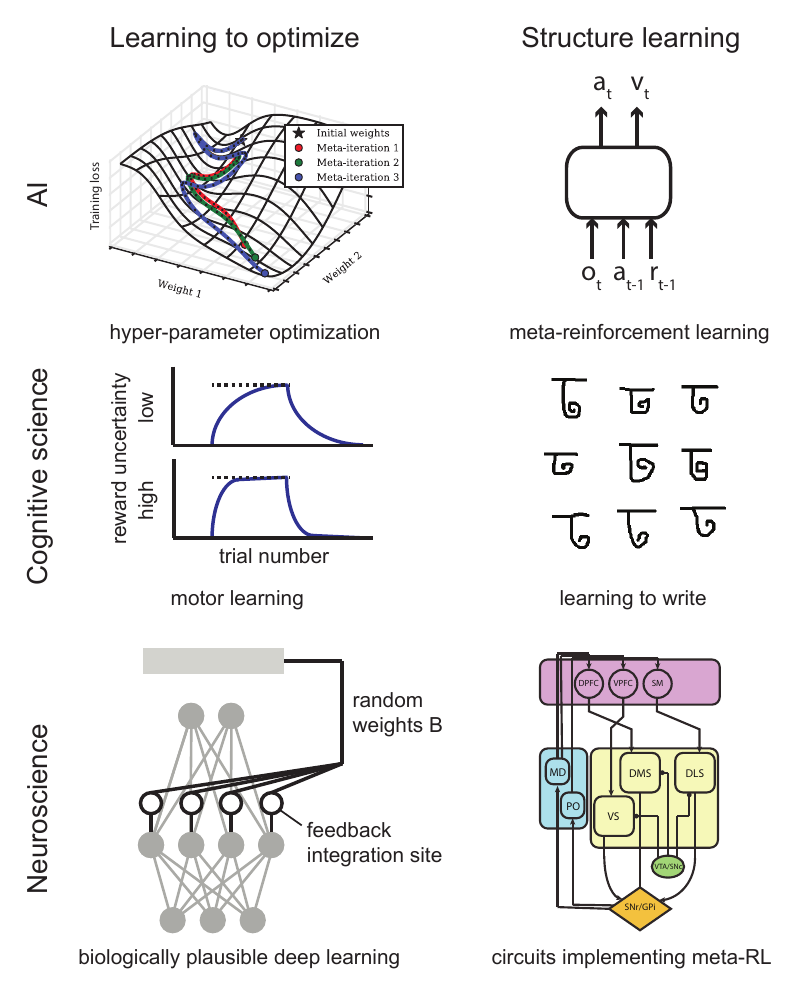}
  \caption{Types of learning-to-learn in AI, cognitive science and neuroscience. Learning-to-learn can be roughly divided into learning to optimize and structure learning. In AI, hyperparameter optimization is an example of learning to optimize \cite{Maclaurin2015-kl}[“Gradient-based Hyperparameter Optimization through Reversible Learning.” Maclaurin D, Duvenaud D, Adams R; PMLR 37:2113-2122 2015. CC BY 4.0], while a recurrent neural network taking rewards, actions and observations can often be used to perform structure learning \cite{Wang2018-di}. In cognitive science, learning rates in motor learning tasks are modulated by uncertainty in the environment -- humans learn faster when it is advantageous to do so \cite{Turnham2012-pn}, while learning to write novel characters can be modeled by a Bayesian Program learning framework \cite{Lake2015-qt} (From “Human-level concept learning through probabilistic program induction.” Lake BM, Salakhutdinov R, Tenenbaum JB, Science 350 (6266) 2015]. Reprinted with permission from AAAS). In neuroscience, biologically plausible models of deep learning, including compartmental models with random feedback weights, produce learning-to-learn effects \cite{Lillicrap2016-qy, Guerguiev2017-lp}, and models of prefrontal cortical circuits have been proposed to form a meta-reinforcement learning circuit, which can perform structure learning \cite{Wang2018-di}. Reprinted by permission from Springer Nature: Nature Neuroscience. Prefrontal cortex as a meta-reinforcement learning system, Jane X. Wang et al, \textcopyright 2018.}
\end{figure}

Learning that becomes more efficient throughout learning many tasks comes in many flavors and has been called meta-learning, learning-to-learn, or structure learning \cite{Thrun1998-te,Kemp2010-ch,Kemp2008-vp}. Two motivating questions when studying learning-to-learn, that are also relevant in cognitive science and neuroscience, are: what are efficient optimization procedures, and how can an agent quickly generalize to new tasks? Based on these two questions we will describe manifestations of learning-to-learn as either learning to optimize or structure learning (Fig. 1). Learning to optimize is the learning of efficient learning rules that quickly reach good solutions. Learning to optimize may occur in a system that solves many unrelated tasks. Structure learning, on the other hand, occurs when systems learn common structure that generalizes to related tasks, allowing for fast learning on new tasks. Structure learning thus can occur only when the system is confronted with a family of tasks that do indeed share meaningful structure. The distinction is not perfectly clear-cut, as methods that can learn to optimize can sometimes also be used for structure learning if presented with the right data. However the distinction usefully divides the ways in which learning-to-learn leads to efficiently learning.

Learning to optimize can provide efficient learning rules without hand-selection. As early as 1991, scientists have were learning synaptic learning rules in artificial neural networks \cite{Bengio1991-ca}. This idea has recently been explored in a wide range of problems to find efficient learning rules for supervised learning tasks \cite{Andrychowicz2016-fr,Chen2017-rq,Wichrowska2017-ar,Metz2018-uz,Zoph2017-pf,Maclaurin2015-kl,Baker2017-mx}. As an example: a common approach to train a system is to incrementally update its parameters in a direction related to the gradient of the performance, so-called gradient descent. There are a range of optimization methods that use gradients, and which method is best for a given task is generally determined through a mix of expert knowledge, heuristics and experimentation. Within a set of possible learning rules, however, the search for which one reaches a solution the fastest can itself be described as an optimization problem. In this way these systems can learn to optimize \cite{Andrychowicz2016-fr}. Rules learnt in this fashion can be designed to have desirable features -- e.g. to be efficient and scalable \cite{Wichrowska2017-ar}. Learning to optimize can make learning more efficient than hand-tuned rules.

Structure learning allows a learner to quickly learn new tasks by exploiting similar structure. For instance, we can see the effect of structure learning when a system that is trained to identify one animal species can quickly be retrained to identify a different species. This works because common structure can be quickly recognized, for instance through feature detectors of eyes, noses, etc. \cite{Long2018-zm,Donahue2014-mj,Yosinski2014-hm}. This can be used to perform what is called few-shot learning -- learning a novel concept from only a handful of examples \cite{Santoro2016-mj,Rezende2016-po}. In this way structure learning allows better transfer.

An agent that learns to learn can be conceptualized as consisting of two separate systems. One system learns to perform a given task -- a learner -- and a separate system adjusts the learner over the course of learning to be more efficient -- a meta-learner \cite{Wang2016-ux}. This basic two-learner setup encompasses (explicitly or implicitly) many implementations of learning-to-learn.

Methods that learn-to-learn differ in the form of the learner and meta-learner. The learner may be restricted in form, or very general. In synthetic gradient methods \cite{Jaderberg2017}, for instance, the learner is restricted to the same form as a neural network trained through the common backpropagation algorithm. One the other hand, a very general approach is to implement a meta-learner as a recurrent neural network (RNN) that itself implements a learning algorithm of arbitrary form. This RNN estimates the parameters of the learner. Over many tasks the parameters of this RNN are updated, which in turn affects the behavior of the learner. This method has been used in reinforcement learning settings to enable learning that transfers across tasks \cite{Wang2016-ux,Duan2017-gy,Mishra2018-fi,Spector2017-ji,Ritter2018-eq,Ritter2018-mf}. The resulting models show a rich set of policies that adapt to the environment, utilizing model-based or model-free learning as appropriate to the task \cite{Wang2018-di}. Similar approaches have also been used in supervised learning settings for few-shot classification of images \cite{Santoro2016-mj,Ravi2017-wa,Vinyals2016-th}. Recently, many other techniques for learning to learn have been developed (e.g. \cite{Finn2017}). Recent research in AI has explored many approaches to implement learning-to-learn. 

\section{Learning-to-learn in cognitive science}

How humans learn efficiently is of fundamental importance to cognitive science and psychology. Humans are particularly efficient learners, \cite{Lake2017-gi,Braun2010-cz,Duncan1960-hh,Barnett2002-qs,Lake2015-qt}, but learning-to-learn is also exhibited in other animals to some extent \cite{Harlow1949-yj}. 

There is evidence that humans learn how to optimize. For instance, in environments in which fast learning is advantageous, people do learn faster \cite{Burge2008-xe,Turnham2012-pn}. Learning many tasks seems to generically increase the speed of learning \cite{Seidler2004-qb}. This matches our experiences in life: people are usually fast at solving the learning problems that they have to solve frequently. 

Humans clearly also perform structure learning. For example, manipulations to a motor control task that are within a given family (e.g. the input to output mapping is rotated by a variable amount) are learnt more quickly than arbitrary manipulations, suggesting that humans can learn an abstract notion of ‘rotation’ from the family of tasks \cite{Braun2010-cz}. Other examples include causal learning \cite{Kemp2010-ch,Gopnik2004-tj,Griffiths2005-rf,Griffiths2009-ba,Kemp2008-vp}, temporal structure learning \cite{Schapiro2013-qd,Schapiro2016-lh} and grammar learning \cite{Schapiro2015-br,Lieberman2004-yl}. Thus structure learning supports humans’ impressive learning abilities.

At a computational level, structure learning can be conceptualized as the acquisition and use of prior knowledge for a given task and thus it can be understood within a Bayesian framework. Hierarchical Bayesian models exhibit structure learning in a range of settings \cite{Gershman2010-ba,Lake2015-qt,Kemp2010-ch,Kemp2008-vp,Gopnik2014-hq,grant2018recasting}. Bayesian Program Learning, for instance, has recently been used to learn to write novel characters with human-like efficiency and style \cite{Lake2015-qt}. The acquisition and use of prior knowledge is thus one approach to understanding learning-to-learn. 

Cognitive models of learning-to-learn focus on computational-level details of how prior knowledge is brought to bear on a problem, while current approaches in AI focus on data-driven neural networks which start as a blank slate and acquire prior knowledge throughout the learning process. Recently, a popular meta-learning method in AI, MAML \cite{Finn2017}, has been cast as a hierarchical Bayesian method \cite{grant2018recasting}, demonstrating how the two fields can provide different views of the same solution. These viewpoints in part reflect an old tension between computational and connectionist approaches in cognitive science, and progress can come by considering their relative strengths and weaknesses. For instance, soon after a Bayesian model of human handwriting \cite{Lake2015-qt} was shown to realistically mimic human behavior and performance, deep learning researchers tackled the same problem using neural networks and without prescribing a detailed hierarchical structure. These networks were able to perform few-shot learning, matching human-level performance on a simplified version of the handwriting task \cite{Santoro2016-mj}. Though it must be kept in mind that reaching this level of performance still requires a large amount of data to prepare the system for the few-shot learning. Regardless, cognitive models of learning-to-learn have prompted progress in AI.

\section{Learning-to-learn in neuroscience}

While it is clear that people get better at learning if they engage in the process of learning, we know little about the way the brain implements this. However recent work has started to address how learning-to-learn may be implemented by neural circuits. 

There are a number of aspects of neural systems that relate to learning to optimize, even if not explicitly understood as such by neuroscientists. First, the brain is full of learning signals and neuromodulators related to outcomes, rewards and prediction errors which enable and control learning \cite{Doya2002-vu,Schultz2016-ia}. Some neuromodulators are known to affect the speed of learning \cite{Doya2002-vu,Yu2005-kk}. These have recently been proposed to have a role in metaplasticity, explaining how learning can proceed more quickly in environments that change more quickly, thus learning to optimize \cite{Roelfsema2018-ek,Iigaya2016-tg,Farashahi2017-dz}. Second, a form of learning to optimize naturally occurs in the reinforcement learning literature. There, parts of the brain may be seen as an actor-critic pair, where the actor learns according to signals from the critic and the critic learns to represent the signals that enable efficient learning for the actor  -- the critic learns to produce the right learning signal \cite{Sutton1998-aj,Fremaux2013-af,Song2017-qi}. And third, more subtle learning-to-learn signatures are present in recently popular models of neuronally realistic backpropagation. In feedback alignment, for instance, neurons use random projections for error information. Throughout learning the system is shown to improve at learning, as estimated gradients become more similar to the real ones \cite{Lillicrap2016-qy}. Moreover, it has been proposed that parts of individual pyramidal cells may learn to represent learning signals for the rest of the neuron \cite{Guerguiev2017-lp,Kording2000-zr}. There is some exploration of learning to optimize across the neuroscience literature, but the potential computational role of neuromodulators and plasticity rules to modulate learning is far from fully explored.

Recent work has also started to address how structure learning may be implemented by neural circuits. While the implementation of computational-level Bayesian models remain to be elucidated, deep learning with artificial neural networks can provide a more direct model, and may ultimately give rise to behavior seen in higher-level cognitive models. Mechanisms for structure learning in motor cortex have been proposed based on deep neural networks \cite{Weinstein2017-pk}. And a recent proposal has taken inspiration from meta-reinforcement learning frameworks first studied in AI. According to this proposal, the prefrontal cortex (PFC) acts as a meta-reinforcement learning system \cite{Wang2018-di,Genewein2015-fj,Botvinick2015-vs}, in which the PFC and subcortical circuits implement an RNN trained through model-free reinforcement learning using a striatal dopamine signaling reward prediction error \cite{Schultz2016-ia}. This meta-reinforcement learning framework allows for complicated learning algorithms to be learnt, facilitating transfer across tasks, and matching a number of behaviors observed in structure learning. This work promises to account for some aspects of how neural circuits can efficiently learn, yet how other forms of structure learning may be implemented by other brain regions and in other domains remains to be explored.

\section{Conclusion}

In neuroscience, cognitive science and AI, a long standing problem is how brains, minds and machines, respectively, can learn efficiently. In recent years learning-to-learn has emerged as an important topic in each of these fields, suggesting its central role to understanding intelligence. Each field contributes distinctly to our overall understanding of learning-to-learn: cognitive science highlights interesting problems, tests, and strategies for AI; artificial neural networks provide inspiration for models in neuroscience and cognitive science and provide insight by placing biologically plausible algorithms in a wider context; and neuroscience constrains models and can reciprocally provide inspiration for models in AI. While cognitive science and AI share an intertwined history, the relation between neuroscience and AI has recently gained attention, owing largely to the success of deep learning. An integration of these fields is far from complete. However, from models of human handwriting inspiring progress in one-shot learning, to models of meta-reinforcement learning in AI inspiring models of structure learning in neuroscience, it is clear that further interaction between the disciplines will yield important insight. The time is ripe for learning-to-learn to find a central role in all three areas.

\section*{Acknowledgements}

The authors would like to thank Adam Marblestone, David Rolnick and Timothy Lillicrap for helpful feedback and discussion.


\bibliography{references.bib}

\end{document}